\documentclass[aps,prl,twocolumn,showpacs,superscriptaddress,groupedaddress]{revtex4}  
\usepackage[dvips]{graphicx}
\usepackage{dcolumn}   
\usepackage{bm}        
\usepackage{amssymb}   

\begin{document}


\title{Waveform-Controlled Terahertz Radiation from the Air Filament Produced by Few-Cycle Laser Pulses}


\author{Ya Bai, Liwei Song, Rongjie Xu, Chuang Li, Peng Liu}
\email{peng@mail.siom.ac.cn}
\author{\mbox{Zhinan Zeng}, Zongxin Zhang, Haihe Lu, Ruxin Li}
\email{ruxinli@mail.shcnc.ac.cn}
\author{Zhizhan Xu}
\email{zzxu@mail.shcnc.ac.cn}
\affiliation{State Key Laboratory of High Field Laser Physics, Shanghai Institute of Optics and Fine Mechanics, Chinese Academy of Sciences, Shanghai 201800, P. R. China}


\date{\today}

\begin{abstract}
Waveform-controlled Terahertz (THz) radiation is of great importance due to its potential application in THz sensing and coherent control of quantum systems. We demonstrated a novel scheme to generate waveform-controlled THz radiation from air plasma produced when carrier-envelope-phase (CEP) stabilized few-cycle laser pulses undergo filamentation in ambient air. We launched CEP-stabilized 10 fs-long ($\sim 1.7$ optical cycles) laser pulses at 1.8 $\mu$m into air and found that the generated THz waveform can be controlled by varying the filament length and the CEP of driving laser pulses. Calculations using the photocurrent model and including the propagation effects well reproduce the experimental results, and the origins of various phase shifts in the filament are elucidated. 
\end{abstract}


\maketitle



Air-plasma based terahertz (THz) wave generation attracts much attention as it provides unique tools for nonlinear spectroscopy, imaging and remote sensing \cite{Kampfrath,Ulbricht, JLiu}. As an intense femtosecond laser pulse undergoes filamentation in ambient air, axial polarized THz radiation in forward direction was generated due to the transition-Cherenkov-type radiation \cite{DAmico}. More intense THz radiation field (up to MV/cm range) can be produced by using two-color laser field, where four wave mixing model and photocurrent model are used to explain the underlying physics \cite{Xie, Kim, Thomson, Theberge}. The spatiotemporal dynamics of THz emission from the air plasma generated by the two-color laser field has recently been discussed and the scheme of tailoring the THz emission spectrum was proposed by adjusting the tunneling ionization events \cite{Babushkin11, Babushkin10}. 

Alternatively, intense THz radiation can be generated from air plasma driven by an intense few-cycle laser pulse. By measuring the amplitude and polarity of generated THz emission, the carrier-envelope-phase (CEP) of few-cycle laser pulses can be inferred \cite{Kress}. The THz emission was interpreted as a result of asymmetric ionization induced quasi-dc current produced by the few-cycle laser pulse, as confirmed in subsequent theoretical investigations \cite{Silaev, Wu}.

It is well understood that a converging Gaussian beam experiences a phase shift of $\pi$, the so-called Gouy phase shift through the laser focus, which has been measured using the stereo above-threshold ionization (ATI) method in high vacuum conditions \cite{Lindner}.  However, the existence of Gouy phase shift in femtosecond laser filaments in air remains controversial \cite{Chin, YLiu}. 

Recent investigations show that waveform-controlled THz radiation is of great importance due to its potential application in THz imaging and coherent control of molecular dynamics \cite{Chan, Kitano}. Can we control the waveform of femtosecond laser filament based THz radiation? Understanding the laser phase effect and propagation effects on the THz radiation from the few-cycle laser pulse produced filament is crucial for the THz generation based CEP metrology \cite{Kress} and the generation of intense and waveform-controlled THz radiation.

In this Letter we propose and demonstrate a novel scheme to generate waveform-controlled THz radiation from the air filament produced by few-cycle laser pulses. Variation of THz waveform and even its polarity inversion are found in the spatially-resolved measurement of THz emission along the filament. THz waveform can be controlled by varying the filament length and the initial CEP of the driving laser pulses.

\begin{figure}[t!]
 \includegraphics{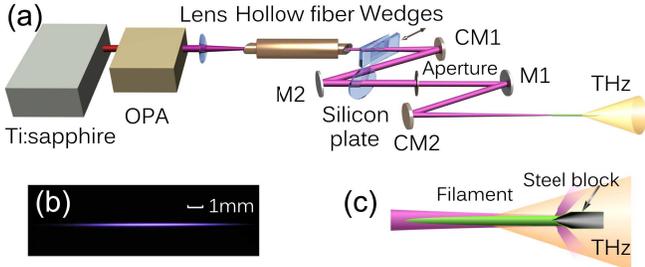}
 \caption{\label{fig:1} (a) Schematics of the experimental layout; (b) Image of the filament formed by focusing the few-cycle laser pulse into ambient air; (c) Schematics of the filament length control.}
 \end{figure}

The CEP stabilized infrared (IR) few-cycle laser pulses are produced using a home-built optical parametric amplifier (OPA) system and a hollow fiber compressor, schematically shown in Fig.~\ref{fig:1}(a). The OPA and compressor are in principle similar to our previous system \cite{Li} except that the pumping laser is the Ti:sapphire laser amplifier facility (Coherent Elite-HP-USX) with pulse energy of 5.2 mJ, duration of 25 fs and repetition rate of 1 kHz at 800 nm. The major portion of the output beam, 4.6 mJ, is used to pump the OPA system while the left as a probe pulse for sampling the generated THz temporal waveforms. In the three-stage OPA amplifier, the idler beam of the $3^{rd}$ stage OPA amplifier is used as the output, so its CEP is passively stabilized with a rms jitter of $\sim$ 400 mrad. The 40 fs-long CEP-stabilized laser pulses at 1.8 $\mu$m are then compressed down to 10 fs ($\sim$ 1.7 cycles) with the maximum pulse energy of 450 $\mu$J. When the laser pulse is focused into ambient air by using a spherical mirror f = 150 mm, a stable luminescence filament of $\sim$ 12 mm-long is formed by using 300 $\mu$J pulse energy selected by an iris diaphragm, as shown in Fig.~\ref{fig:1}(b). The generated THz radiation are measured by using the standard balanced diode geometry electric-optic (EO) sampling technique.

In order to investigate the evolution of driving laser pulse and THz radiation in the filament, we insert a sharp stainless steel blade ($\sim$ 0.2 mm $\times$ 4 mm $\times$ 20 mm) into the plasma column to stop the filament, which can be moved by a motor stage along the laser propagation direction defined as the $z$ coordinate henceforth. As shown in Fig.~\ref{fig:1}(c), the blade is positioned to allow the laser beam hit onto its sharp edge, so that the detection of THz radiation from the left filament is lest influenced. In so doing, the integrated THz signals from filaments of different lengths are measured. We calibrate experimentally the deviation in the collection efficiency of the THz signal detection optics due to the change in the distance between the filament and the detector.

\begin{figure}[b!]
\includegraphics{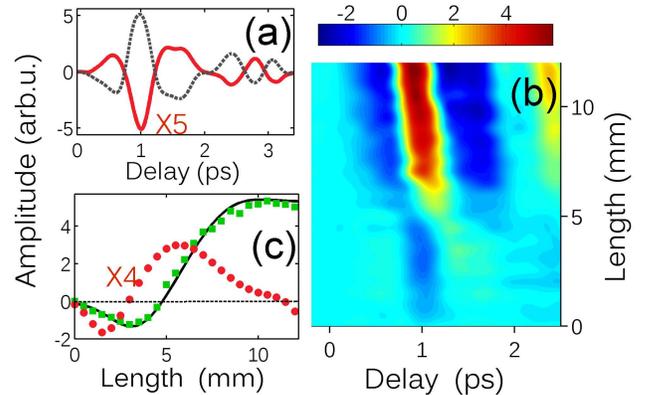}
 \caption{\label{fig:2} (a) THz temporal waveforms measured from the filaments of different length 3 mm (red solid line) and 10 mm (black dashed line), respectively; (b) THz waveforms measured by moving the block continuously along the filament; (c) The measured THz amplitudes (green solid squares) taken from Fig. 2(b) at the delay time of 1 ps, the calculated results (black solid line), and the retrieved THz amplitude (red solid circles) as a function of the position in the filament.}
 \end{figure}

Fig.~\ref{fig:2}(a) shows two THz waveforms by blocking the filament at the position of 3 mm (the distance from the visible starting position of the filament to the blade) and 10 mm, respectively. It is reasonable that the measured THz amplitude from the 3 mm-long filament is much smaller than from the 10 mm-long one, since the measured signal comes from the integrated emission from the unblocked filament. One can also note that the recorded THz radiation from the 10 mm-long filament reverses its polarity comparing to that from the 3 mm-long filament, indicating the phase variation of the driving laser field in the filament. By moving the blocking blade continuously in a step length of 0.5 mm, the THz waveforms as a function of filament length are recorded and plotted in Fig.~\ref{fig:2}(b). One can see that, the THz emission gradually changes its amplitude and polarity when the length of filament is varied, and at certain length ($\sim$ 5 mm-long filament) the THz emission signal goes to zero and then reverses the polarity thereafter. One can therefore control the THz waveform by varying the filament length.

For analyzing the amplitude modulation of THz radiation, the amplitude values when the THz signal is the maximum, at $\sim$ 1 ps in Fig.~\ref{fig:2}(b), are plotted in Fig.~\ref{fig:2}(c). With the help of the above-mentioned calibration of collection efficiency, the spatially resolved THz emission as a function of the position in the filament can be retrieved, also shown in Fig.~\ref{fig:2}(c). One can see that the THz emission first increases in one (negative) polarity and then decreases to zero at $\sim$ 3 mm, after which the amplitude increases in the other (positive) polarity and decreases at the tail of the filament. 

The polarization of THz emission was measured to be nearly linear so that we excludes that the generated THz emission is originated from transition-Cherenkov-type radiation. We consider the THz emission originated from the transient photocurrent driven by the propagating intense few-cycle laser fields in the plasma \cite{Kim}. The propagation of intense few-cycle laser fields in a dispersive medium can be described by using a propagation equation in an axial-symmetric coordinates \cite{JSLiu},
\begin{eqnarray}\label{eq:1}
\partial_{z}\widetilde{E}(\mbox{\boldmath $r$},z,\omega)
= [\frac{i}{2k(\omega)}\bigtriangledown_{\perp}^{2}+ik(\omega)]\widetilde{E}(\mbox{\boldmath $r$},z,\omega)
\nonumber\\
+ \frac{i\omega^{2}}{2c^{2}\varepsilon_{0}k(\omega)}\widetilde{P}_{NL}
- \frac{\omega}{2c^{2}\varepsilon_{0}k(\omega)}\widetilde{J}_{ioni}
\end{eqnarray}
where $\widetilde{E}(\mbox{\boldmath $r$},z,\omega)$  is the frequency domain laser field, the nonlinear polarization $\widetilde{P}_{NL}$  accounts for the Kerr effect and the polarization $\widetilde{J}_{ioni}$ is caused by photoelectrons from the tunneling ionization of $N_{2}$ and $O_{2}$ in air. The collective motion of the tunneling ionized electrons results in a directional nonlinear photocurrent surge described by \cite{{Babushkin10}}
\begin{equation} \label{eq:2}
\partial_{t}J_{e}(\mbox{\boldmath $r$},z,t)+\nu_{e}J_{e}(\mbox{\boldmath $r$},z,t)
=\frac{e^2}{m}\rho_{e}(\mbox{\boldmath $r$},z,t)E(\mbox{\boldmath $r$},z,t)
\end{equation}
where $\nu_{e}, e, m$ and $\rho_{e}$ denote the electron-ion collision rate, electron charge, mass, and electron density, respectively. The transient current at each propagation step of the calculation is treated as radiation source of far field THz emission \cite{Kohler}
\begin{equation} \label{eq:3}
E_{THz}(\mbox{\boldmath $r'$},t)=-\frac{1}{4\pi\epsilon_{0}}\int \frac{1}{c^2R}\partial_{t}J_{e}(\mbox{\boldmath $r$},z,t_{r})d^3\mbox{\boldmath $r$}
\end{equation}
where $R$ is the distance between the THz point source and the detection plane. Eq.~(\ref{eq:3}) describes the THz radiation towards all directions. The forward far field THz emission near the propagation axis from the unblocked air plasma has been added up to simulate the measured results.  The initial few-cycle laser field is of a Gaussian beam with the waist size of $w_{0}$ = 260 $\mu$m ($1/e^2$) at center wavelength of 1.8 $\mu$m, duration of $\tau$ = 10 fs in full width at half maximum (FWHM) and pulse energy of 300 $\mu$J. By optimizing the initial CEP of the driving few-cycle laser field to be $\varphi_{0}$ = 0.33 $\pi$, the THz amplitude as a function of filament length is found in consistence with the measured results, as shown in black solid line in Fig.~\ref{fig:2}(c).

 \begin{figure}[t!]
\includegraphics[width=0.5\textwidth]{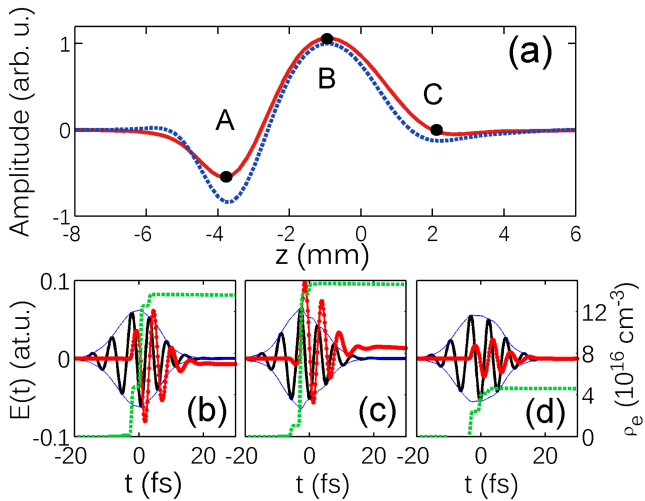}
 \caption{\label{fig:3}(a) The calculated THz amplitude (red solid line) and asymmetry parameter (blue dashed line), $\sum(J_{e}^+ - J_{e}^-)$, in air plasma filament; (b-d) At positions of A, B and C, the laser field envelopes (blue thin line), carrier oscillations (black thick line), electron densities (green dashed line) and transient photocurrents (red dotted solid line).}
 \end{figure}


For a further understanding of the THz emission characteristics in the air plasma, we show in Fig.~\ref{fig:3}(a) the space-resolved THz emission calculated from Eq.~(\ref{eq:3}) along the filament. As one can see that the THz amplitude modulates from negative polarity at the beginning of plasma to positive one at the end, which results in the inversed THz emission polarity shown in Fig.~\ref{fig:2}(c). In a simplified picture, the d.c. transverse current is related to discrete electron ionization events and the pulse vector potential $A(t_{i})$ modulation \cite{Babushkin10, Geissler}, which is described by $J_{e}\propto \sum_{i}\delta\rho_{i}A(t_{i})$. In order to distinguish electrons with opposite drifting velocities, we define two quantities  $J_{e}^+$ and $J_{e}^-$, which describe the positive current density and negative current density, respectively. We calculate the asymmetry of the summed current densities $\sum(J_{e}^+ - J_{e}^-)$ which matches the THz emission as shown in Fig.~\ref{fig:3}(a).

The laser fields, electron densities and transient photocurrents at three positions labeled as A, B and C in Fig.~\ref{fig:3}(a) are calculated and plotted in Figs.~\ref{fig:3}(b)--~\ref{fig:3}(d), respectively. At the position of A, where the THz emission is the maximum in the negative polarity, the amplitude of negatively current density  $J_{e}^-$  is larger than the positively current density  $J_{e}^+$ following the laser field oscillation. At the position of B, the negative current density is less than the positive one so that the THz radiation is positively polarized. At the position of C, the almost equal current densities in the oscillation result in the near zero THz radiation. From Figs.~\ref{fig:3}(b)--~\ref{fig:3}(d), one can also note that the increments of produced electron density $\rho_{e}$ are mainly from two attosecond bursts in the opposite directions, that are originated from the tunneling ionization in a single optical cycle, making the THz emission extremely sensitive to the variation of driving laser fields. The attosecond electron bursts can be easily controlled by varying the intense few-cycle fields, comparing to the THz radiation generation by two-color multi-cycle laser fields \cite{Babushkin11}.

Since the THz emission is shown to be sensitively determined by the intense few-cycle laser field, we then investigate the CEP variation in the air plasma. As indicated previously \cite{{Porras}}, the CEP is the difference of the pulse phase, i.e. carrier phase, and the pulse front. The carrier phase shift in a dispersive medium can be written as
\begin{equation} \label{eq:4}
\phi_{\omega_{0}}(\mbox{\boldmath $r$},z)=n(\mbox{\boldmath $r$},z) \frac{\omega_{0}}{c}z
+\phi_{G} (z,z_{eff}) + \phi_{off-a}(\mbox{\boldmath $r$},z)
\end{equation}
where the first term is the phase shift at the carrier frequency induced by spatial dependent refractive index $n($\mbox{\boldmath $r$}$,z)$, the second term is the Gouy phase shift with the effective focal length $z_{eff}$, and the third term denotes the off-axial phase shift due to diffraction effects. For simplicity we only consider the axial phase shift that includes the first two terms. We define the second term as the modified Gouy phase shift that is characterized by $z_{eff}$, since the Rayleigh length is no longer valid in filament.

 \begin{figure}[t!]
\includegraphics[width=0.5\textwidth]{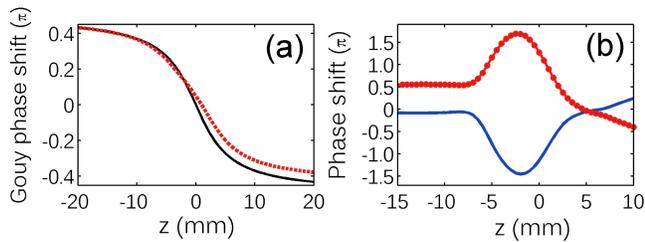}
 \caption{\label{fig:4}(a) The modified Gouy phase shift (red dashed line) and the Gouy phase shift (black solid line) of linear focus condition; (b) The pulse front shift (blue solid line) and the CEP shift (red dotted solid line) of the laser field in air plasma.}
 \end{figure}

The calculated modified Gouy phase shift decreases with a relatively slower slope than the Gouy phase shift under the linear focus condition, as shown in Fig.~\ref{fig:4}(a). The modified Gouy phase shift due to the extension of filament confirms the previous theoretical prediction that the Gouy phase shift stems from transverse spatial confinement of a finite beam \cite{Feng}. The CEP of laser fields is also determined by the propagating pulse front, which is defined as Eq. (17) of the Ref. \cite{Porras} and can be retrieved from the calculation results. The axial pulse front is shown in Fig.~\ref{fig:4}(b). As one can see the pulse front experiences a jump around the center of the plasma region (z = -7 mm $\sim$ 4 mm), as a result of the plasma-induced dispersion. The total axial CEP shift is $\sim$ 1.4 $\pi$ and possessing a hump structure in the plasma region, which is originated from the modified Gouy phase shift and plasma effects.

It should be noted that the modified Gouy phase shift also applies in the filament produced by multi-cycle laser pulses, in which the CEP is not a significant parameter for THz generation. In the recent studies on THz Air-Biased-Coherent-Detection (ABCD) \cite{Zhang, He}, the Gouy phase shift of the THz emission and the weak ultrafast probe pulses around the focus was observed. The present work reveals that, due to the nonlinear interaction in air plasma, the intense laser field experiences not only the modified Gouy phase shift but also the variation of pulse front.

\begin{figure}[t!]
\includegraphics[width=0.5\textwidth]{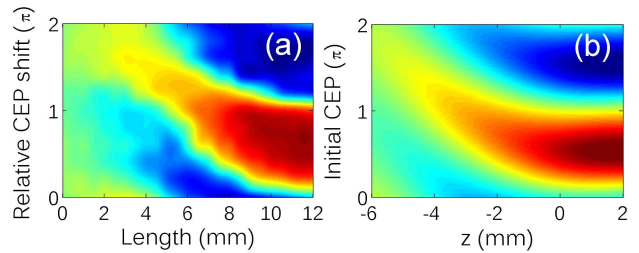}
 \caption{\label{fig:5}(a) Measured and (b) simulated THz amplitude as functions of relative CEP shift and filament length.}
 \end{figure}

Furthermore, we look into the initial CEP, $\varphi_{0}$, dependence of the THz emission from the air plasma. $\varphi_{0}$ is adjusted by passing the laser beam through a pair of thin wedges mounted on a motor stage. By increasing the initial CEP gradually in a step size of 0.2$\pi$, we obtained a series of 2-dimentional THz waveform maps similar to that shown in Fig.~\ref{fig:2}(b). The amplitudes of THz emission along the filament are retrieved and plotted as a function of $\varphi_{0}$ in Fig.~\ref{fig:5}(a). We found that the THz waveforms change back to be the same as those shown in Fig.~\ref{fig:2}(b) after a change of initial CEP by 2$\pi$. This indicates that in air plasma the propagation effect on the CEP variation is independent of the initial CEP of driving laser fields, which is also verified by our numerical simulations. The phases of few-cycle pulses experience a fixed shift of over $\pi$, which is consistent with the CEP shift through the air plasma. This observation helps to validate the CEP metrology based on THz generation \cite{{Kress}}. However, due to the variation of CEP through the plasma, an offset of CEP must be determined and taken into account for measuring the CEP of laser pulses. As shown in Fig.~\ref{fig:5}(b), the simulated results using the photocurrent model and propagation equation agree well with experimental results. Since the polarity of THz radiation can be controlled by varying the initial CEP of driving laser pulses, one can maximize the output THz radiation intensity from the filament by optimizing the initial CEP to minimize the intensity cancellation due to opposite polarities.

In conclusion, we have demonstrated a scheme to generate waveform-controlled THz radiation from air plasma produced by CEP-stabilized few-cycle laser pulses. The generated THz waveform can be controlled by varying the length of the filament  and the CEP of the driving laser pulses. The waveform evolution of THz radiation in the filament is due to the characteristic carrier phase shift and variation of laser envelope during propagation. Due to the ability to control the spatial location of femtosecond laser filamentation using pre-chirped laser pulses, the demonstrated scheme of waveform-controlled THz radiation generation is suitable for remote sensing exploring the phase property of THz radiation. The knowledge about the phase shift in a filament would also be valuable for optimizing the generation of phase-matched attosecond pulses driven by intense few-cycle laser pulses.

\begin{acknowledgments}
This work is supported by Chinese Academy of Sciences, Chinese Ministry of Science and Technology, NNSF of China (Grant Nos. 60921004, 10734080, 10523003, 60978012, and 11134010), 973 Program of China (2011CB808103), and Shanghai Commission of Science and Technology (No. 08PJ14102). 
\end{acknowledgments}


\end{document}